# Disentangling Spatial-Temporal Functional Brain Networks via Twin-Transformers

Xiaowei Yu[1], Lu Zhang[1], Lin Zhao[2], Yanjun Lyu[1], Tianming Liu[2], Dajiang Zhu[1]

[1] Department of Computer Science, University of Texas at Arlington

[2] Department of Computer Science, University of Georgia

**Abstract.** How to identify and characterize functional brain networks (BN) is fundamental to gain system-level insights into the mechanisms of brain organizational architecture. Current functional magnetic resonance (fMRI) analysis highly relies on prior knowledge of specific patterns in either spatial (e.g., resting-state network) or temporal (e.g., task stimulus) domain. In addition, most approaches aim to find group-wise common functional networks, individual-specific functional networks have been rarely studied. In this work, we propose a novel Twin-Transformers framework to simultaneously infer common and individual functional networks in both spatial and temporal space, in a self-supervised manner. The first transformer takes space-divided information as input and generates spatial features, while the second transformer takes time-related information as input and outputs temporal features. The spatial and temporal features are further separated into common and individual ones via interactions (weights sharing) and constraints between the two transformers. We applied our Twin-Transformers to Human Connectome Project (HCP) motor task-fMRI dataset and identified multiple common brain networks, including both task-related and resting-state networks (e.g., default mode network). Interestingly, we also successfully recovered a set of individual-specific networks that are not related to task stimulus and only exist at the individual level.

## 1 Introduction

Using functional magnetic resonance imaging (fMRI) to reconstruct concurrent brain networks has been of intense interest in neuroscience for years [1, 2]. The brain networks not only provide spatial and temporal information of the brain, but they also have clinical potentials as non-invasive imaging biomarkers. To date, it has been proven that the task-activated brain networks (BNs) can be reconstructed from task-fMRI while the resting-state BNs can be obtained from resting-state fMRI (rs-fMRI) [3, 4]. Recently, some studies have shown that the brain networks, including task-activated and resting-state BNs, can be inferred from task-fMRI simultaneously [5, 6]. Independent component analysis (ICA) is one of the most popular ways to identify the resting-state brain networks. However, ICA and its variations can be limited in characterizing the FNs with both spatial and temporal overlaps [7]. General linear models (GLM) are the mainstream methods for task-activated brain networks, but it heavily relies on the prior knowledge of the task design [8]. Sparse learning methods achieve great performance in identifying both task-activated and resting-state BNs, and the corresponding temporal patterns of these BNs [9]. However, sparse learning, like other shallow models, can only capture simple relations between spatial and temporal domains. Recent advances in deep learning methods have shed light on addressing these limitations. Numerous deep learning models have been proposed, such as Vision Transformer (ViT) and masked auto-encoder in the computer vision domain, which have shown the versatility of self-attention-based methods in processing images and videos [10, 11]. Due to the fundamental difference between task-fMRI and videos, existing self-attention architectures cannot be directly applied to task fMRI data, which makes it difficult to adopt transformers in task-fMRI applications [12-14]. To the best of our knowledge, there have been few attempts applying self-attention models to 4D task-fMRI data. In general, current methods either use a CNN kernel to preprocess the task-fMRI data and feed pre-processed features into transformers, or use a pre-trained transformer model [15, 16].

To fully take advantage of self-attention models in task-fMRI, we propose a spatial-temporal disentangled twin-transformers network for exploring task-fMRI. The architecture of the proposed model is shown in Fig. 1(a). Due to the spatial-temporal entangled nature of the task-fMRI, we need to consider spatial and temporal information simultaneously. We extract the brain signals from each voxel and organized them into a 2D signal matrix. The signal matrix can be further disentangled into temporal and spatial features. These brain signal matrices are organized into a pair-wise manner to learn the common patterns as well as to enlarge the datasets. Upon the pair-wise input, a spatial-temporal disentangled Twin-Transformers model is proposed, where one transformer is trained to capture common and individual spatial patterns, and the other is trained to learn the common and individual temporal patterns. We evaluated the proposed Twin-Transformers using Human Connectome Project (HCP) motor task-fMRI dataset and identified multiple common brain networks, including both task-related and resting-state networks (e.g., default mode network). We also successfully recovered a set of individual-specific networks that are not related to task stimulus and only exist at the individual level. In this work, our contributions have three folds: 1) We consider the entangled nature of spatial and temporal information in task-fMRI data and propose a spatial-temporal disentangled Twin-Transformers network for brain network discovery using task-fMRI data; 2) This work is one of the earliest works that introduce the transformers into high dimensional fMRI imaging data instead of the extracted features; 3) The proposed Twin-Transformers can identify common brain networks, including both task-related and resting-state networks. Moreover, a set of individual-specific networks are also recovered.

## 2 Methods

### 2.1 Spatial-Temporal Data Preparation

In our experiment, we used task-fMRI data of 917 subjects from the HCP 3T motor task dataset [17, 18]. The task-fMRI images are aligned in MNI space and images downsampled from 2mm to 4mm to reduce computational complexity. We apply the T1 mask of the cerebral cortex to the task-fMRI images and extract 28549 voxels. Each voxel contains a series of brain signals with a length of 284. These signals are treated as column vectors and organized into a 2D matrix. In this way, a 4D task-fMRI imaging can be represented by a 2D matrix, where the rows represent the temporal information, and the columns represent the spatial information (dark blue boxes at the top of Fig. 1 (b)). We normalized the brain signals to zero mean and unit variance. To facilitate patch partitions, we expand the space dimension to 28800 by adding zero vectors along the spatial dimension. Finally, a set of 2D brain signal matrices with dimensions of 284×28800 are generated.

### 2.2 Twin-Transformers

To disentangle the spatial and temporal information contained in the generated brain signal matrices, a Twin-Transformers mode is proposed. The architecture of the Twin-Transformers is illustrated in Fig. 1 (b). There are two transformer components in Twin-Transformers to separately learn spatial and temporal features by focusing on the different dimensions of the input brain signal matrix.

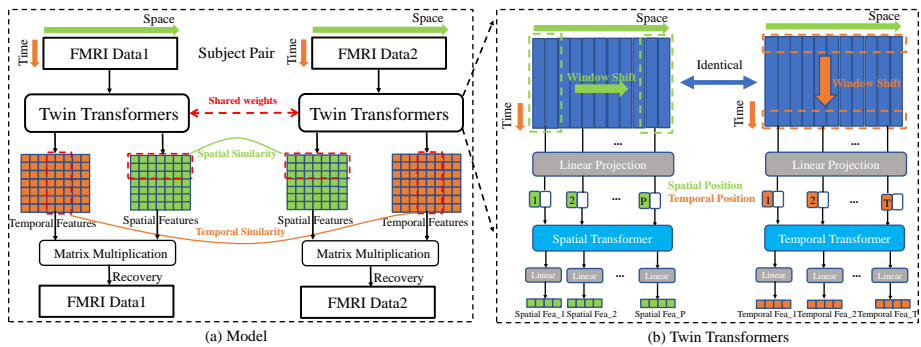

**Fig. 1.** Illustration of the proposed Twin-Transformers framework. (a) shows the overall picture of the proposed model, and (b) shows the details of the Twin-Transformers. The Twin-Transformers take the identical brain signal matrix as input. The spatial Transformer focuses on the space dimension and takes non-overlapping spatial patches as tokens to build attention in the spatial dimension and generate spatial features. Conversely, the temporal Transformer focuses on the temporal dimension and the non-overlapping temporal patches are used as tokens. Correspondingly, the temporal Transformer is designed to build attention in the time dimension and generate temporal features. The twin Transformers are designed for disentangling temporal and spatial features of the input signal matrix.

Specifically, the spatial transformer is designed to learn the latent representations of spatial features. It divides the input signal matrix into $P$ non-overlapping patches by shifting the sliding window (green dotted box) from left to right along the space dimension. The size of the sliding window can be adjusted according to the size of the input data. Each spatial patch contains complete temporal information of the focal brain region. The $P$ patches correspond to $P$ components of brain networks as predefined. During the training process, patches are used as tokens. Each token is first fed into a linear projection layer to get the representation $z_i \in R^{1 \times D_1}$, and then the learnable spatial positional embedding, $E_i^s \in R^{1 \times D_1}$ are added to the representations of each input token. The spatial transformer encoder can be formulated as:

$$Spa(Z) = MLP(MSA(z_1^s || z_2^s || z_3^s || ... || z_P^s)) \qquad (1)$$

where $MSA()$ is the multi-head self-attention, $MLP()$ represents multilayer perceptron, $z_i^s = (z_i + E_i^s), i = 1, 2, ..., P$, and $||$ denotes the stack operation. $Spa(Z) \in P \times N$ is the output of the spatial Transformer, where $P$ represents the number of brain networks and $N$ is the number of voxels in the brain. $Spa(Z)$ models the activated voxels within each brain network.

The temporal transformer is designed to learn the latent representations of temporal features. Similar to the spatial transformer, by shifting the sliding window (orange dotted box) from top to bottom along the time dimension, $T$ non-overlapping temporal patches are generated. The size of the sliding window equals 1, hence the number of patches equals the length of the brain signals. Each temporal patch contains information of all the voxels. After input embedding and positional embedding, each patch is represented by $z_i^t = z_i + E_i^t, i = 1, 2, ..., T$. The temporal transformer encoder can be formulated as:

$$Tem(Z) = MLP(MSA(z_1^t || z_2^t || z_3^t || ... || z_T^t)) \qquad (2)$$

The outputs $Tem(Z)$ of the temporal transformer have a dimension of $Tem(Z) \in T \times P$, where $T$ eques to the time points of the fMRI signals. $Tem(Z)$ represents the signal pattern of each brain network. Taking $Spa(Z)$ and $Tem(Z)$ together, we can obtain both the spatial and temporal patterns of each brain network.

**Spatial-Temporal Commonality-Variability Disentangled Loss**

To simultaneously capture common and individual patterns in the spatial and temporal domain, a new spatial-temporal commonality-variability disentangled loss (ST-CV Loss) is proposed. There are three components in ST-CV Loss. The first one is the signal matrix reconstruction loss. The whole framework is trained in a self-supervised manner to reconstruct the input signal matrix from the learned spatial and temporal features. This is crucial to ensure the learned spatial and temporal features have captured the complete spatial and temporal information of the input data. The reconstruction loss can be formulated as:

$$L_{reco} = \sum \|X - Spa(Z) \cdot Tem(Z)\|_{L1} \qquad (3)$$

where $X$ is the input signal matrix, and we use L1-norm to constrain the reconstruction of the input subject pair. The second component is the commonality constrain loss of spatial features, which aims to disentangle the common and individual spatial features. For this purpose, the learned spatial feature matrix is divided into common part (the first $p$ rows) and individual part (the remaining rows). The common and individual features can be learned by minimizing the difference between common parts of different subjects and leaving the individual parts to learn freely. This can be formulated as:

$$L_{comm\_spa} = \|Spa(\mathbf{Z}_1)[0:p,*] - Spa(\mathbf{Z}_2)[0:p,*]\|_{L1} \qquad (4)$$

where $[0:p,*]$ represents the first $p$ rows in $Spa(\mathbf{Z}_i)$, and $*$ means for each row, all the elements in the columns are included, and vice versa. We adopt the L1 norm to constrain the distance of common spatial features between different subjects to be minimized. Similarly, the commonality constraint on temporal features which is the third component in ST-CV Loss can be formulated as:

$$L_{comm\_tem} = \sum Corr(Tem(\mathbf{Z}_1)[*,p:2p], \ Tem(\mathbf{Z}_2)[*,p:2p]) \qquad (5)$$

Pearson's correlation coefficient calculator – $Corr(\cdot)$ is used to constrain the similarity of common temporal features of different subjects which needs to be maximized. Combining the three parts, the ST-CV Loss can be formulated as:

$$ST\text{-}CV\_Loss = \alpha L_{reco} + \beta L_{comm\_spa} - \gamma L_{comm\_tem} \qquad (6)$$

where the regularization parameters $\alpha$, $\beta$, and $\gamma$.

## 3  Experiments

### 3.1  Experiment Setting

For the spatial transformer, the window size is set to be 288, so the brain signal matrix with dimensions 284×28800 (temporal×spatial) is divided into 100 patches with dimensions of 284×288. According to formula (1), the output size of the spatial transformer is $100 \times 28800$. For the temporal transformer, the window size is set to 1, therefore, 284 patches with dimension 1×28800 are generated. And based on formula (2), the output size of the temporal transformer is $284 \times 100$. For both spatial and temporal transformers, the depth of the transformer is 6, and the multi-head number is 4. For both spatial and temporal transformers, the embedding layer and the fully connected feed-forward network produce outputs of dimension 1024 and 2048, respectively. We adopt the Adam optimizer with a learning rate of 0.001.

### 3.2  Spatial-Temporal Disentangled Common Networks

**Task-related Networks within Common Networks.** After the model is well-trained, we can obtain a set of brain networks, whose spatial and temporal patterns are modeled by $Spa(\mathbf{Z})$ and $Tem(\mathbf{Z})$ (formula (1) and (2)). Brain networks (BNs) whose temporal features follow the task design are recognized as task-related brain networks. We found three common task-related BNs (tongue, right foot (RF), and left foot (LF)) at the common part in $Spa(\mathbf{Z})$ and $Tem(\mathbf{Z})$, and we showed them using 5 randomly selected subjects in Fig. 2. For each subject, the three BNs are shown in the same column at different rows. For each BN, its temporal pattern is displayed at the top and the spatial pattern is shown by the T1 slices on the bottom. As shown in Fig. 2, the temporal pattern (red) is highly correlated to the task design (white), and the corresponding spatial pattern shows the brain regions that are activated in the tasks [16, 17]. For each task-related BN, the activated brain regions can be consistently found in all the subjects. This result suggests that our model can successfully identify the task-related brain networks without any prior knowledge.

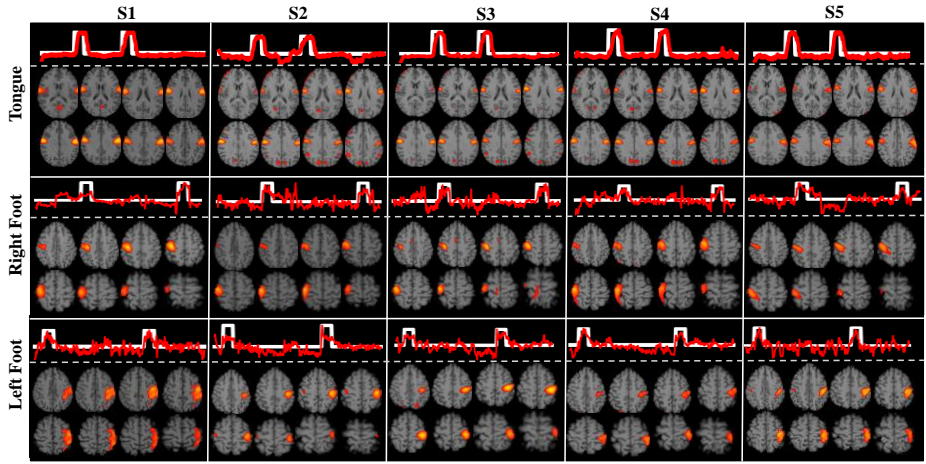

**Fig. 2.** Illustration of task-related brain networks of 5 randomly selected subjects, marked as S1-S5. For each subject, the spatial and temporal patterns of three identified task-related brain networks are displayed (Tongue, Right Foot (RF), Left Foot (LF)). The task designs are shown in white color, while the temporal patterns are shown in red color. The spatial patterns are shown using 8 T1 slices.

**Resting-state Networks within Common Networks.** We identify resting-state BNs by comparing spatial patterns with the template of well-known resting-state networks [3], and 9 common resting-state BNs are recognized. Due to the limited page space, we present 4 of them in Fig. 3 and the remaining 5 can be found in the supplementary. We show the spatial and temporal patterns of the 4 resting-state BNs in 10 randomly selected subjects at the first 10 columns. The template and average spatial pattern are shown in the last two columns. We can see that the spatial pattern of each resting-state BN is consistent among different subjects and have high similarity with the template. Moreover, the BNs in the first row and the fourth row are located in the occipital lobe, which is responsible for vision [19, 20]. This is consistent with the fact that the subjects attempting the experiments are instructed by visual cues. Besides, the BN2 is the default mode network and its temporal patterns are inclined to be anti-task [21, 22], which is consistent with previous studies. The BNs in the third row are mainly located in the frontal lobe and related to motor function [23, 24]. In general, the spatial pattern of the common resting-state brain networks identified by our model shows high similarity with the template and the corresponding temporal patterns of them are consistent with existing research results.

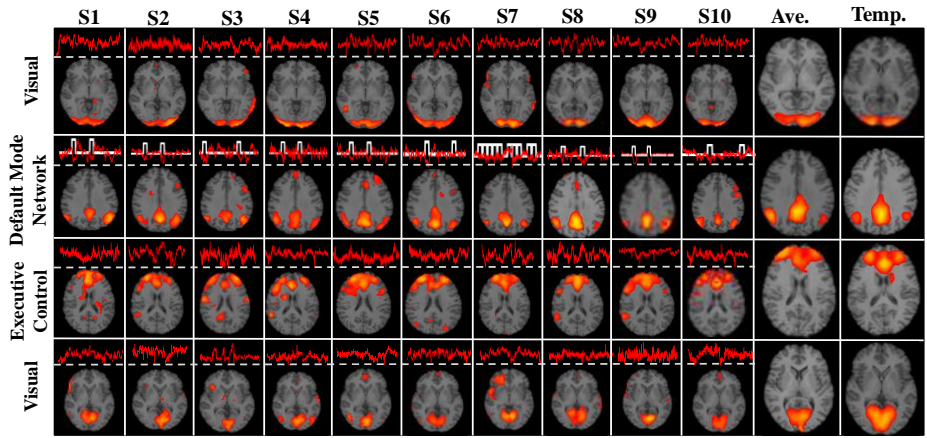

**Fig. 3.** Illustration of resting-state brain networks in 10 randomly selected subjects, marked as S1-S10. For each individual, we show the spatial and temporal patterns of 4 typical brain networks. The last two columns are the average spatial pattern and the well-recognized template.

**Visualization of Individual BNs.** We also found some brain networks that their temporal features do not follow the task design, and their spatial patterns are not consistent with the template. That is, these BNs only exist at the individual level and we recognized them as individual networks. Fig. 4 shows these individual networks using nine randomly selected subjects, where each subject includes three individual brain networks. The individual brain networks indicate that when launching the same task, besides the common brain networks across the subjects, different subjects have their unique brain activities. The existence of individual BNs may be related to the widely existing individual functional variability.

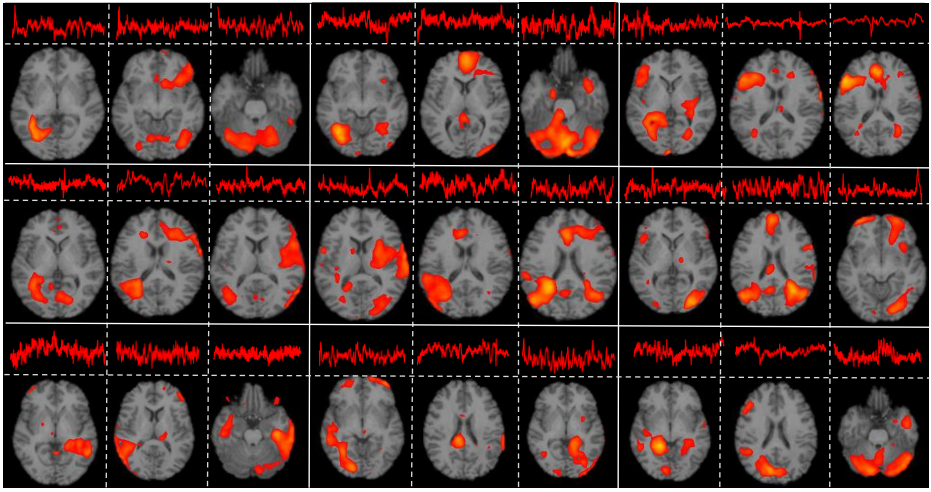

**Fig. 4.** Illustration of individual patterns in nine subjects. The spatial and temporal patterns that emerged from the individual features are various among the subjects.

### 3.3 Reproducibility

To verify the reproducibility of our proposed Twin-Transformers in various parameters settings, for example, different number of components, various common components ratios (CCR), extensive experiments are implemented to test the robustness and stability of the model. Through all experiments, we count the number of task-related BNs and resting-state BNs across all the subjects and calculate the averaged spatial overlaps between resting-state BNs and templates as an index of the performance. We use the Jaccard similarity coefficient [25] to calculate the overlap, which is formulated as below:

$$J = \frac{Template_{BNs} \cap SubjectBNs}{Template_{BNs} \cup SubjectBNs} \qquad (7)$$

Specifically, a larger/smaller $J$ means that the BN is more/less similar to the template. We also measure the PCC between the averaged task-related temporal patterns and task designs. Table 1 shows the performance of different model settings. We can see that the number of task-related and resting-state BNs are stable across different experiments, which indicates that the proposed model can stably and robustly disentangle temporal and spatial features under different settings.

**Table 1.** Experiment results under different settings.

| Comp. | CCR | Task-relate BNs | PCC | Resting-state BNs | $J$ |
|---|---|---|---|---|---|
| 100 | 40% | Tongue, RF, LF | 0.68 | 9 rs-BNs | 0.81 |
| 100 | 60% | Tongue, RF, LF | 0.68 | 9 rs-BNs | 0.80 |
| 200 | 50% | Tongue, RF, LF | 0.64 | 9 rs-BNs | 0.77 |
| 300 | 50% | Tongue, RF, LF | 0.61 | 9 rs-BNs | 0.74 |

## 4      Conclusions

In this paper, we proposed a Twin-Transformers model for brain network discovery. Considering the spatial-temporal entangled property of the task-fMRI, a spatial transformer and a temporal transformer are used to model the spatial and temporal features separately. Moreover, an ST-CV Loss is designed to capture the common and individual patterns simultaneously. We applied the proposed Twin-Transformers on the Human Connectome Project (HCP) motor task-fMRI dataset and identified multiple common brain networks, including both task-related and resting-state networks. Besides, we also recovered a set of individual-specific networks neither related to task stimulus nor consistent with the template.

# Supplementary

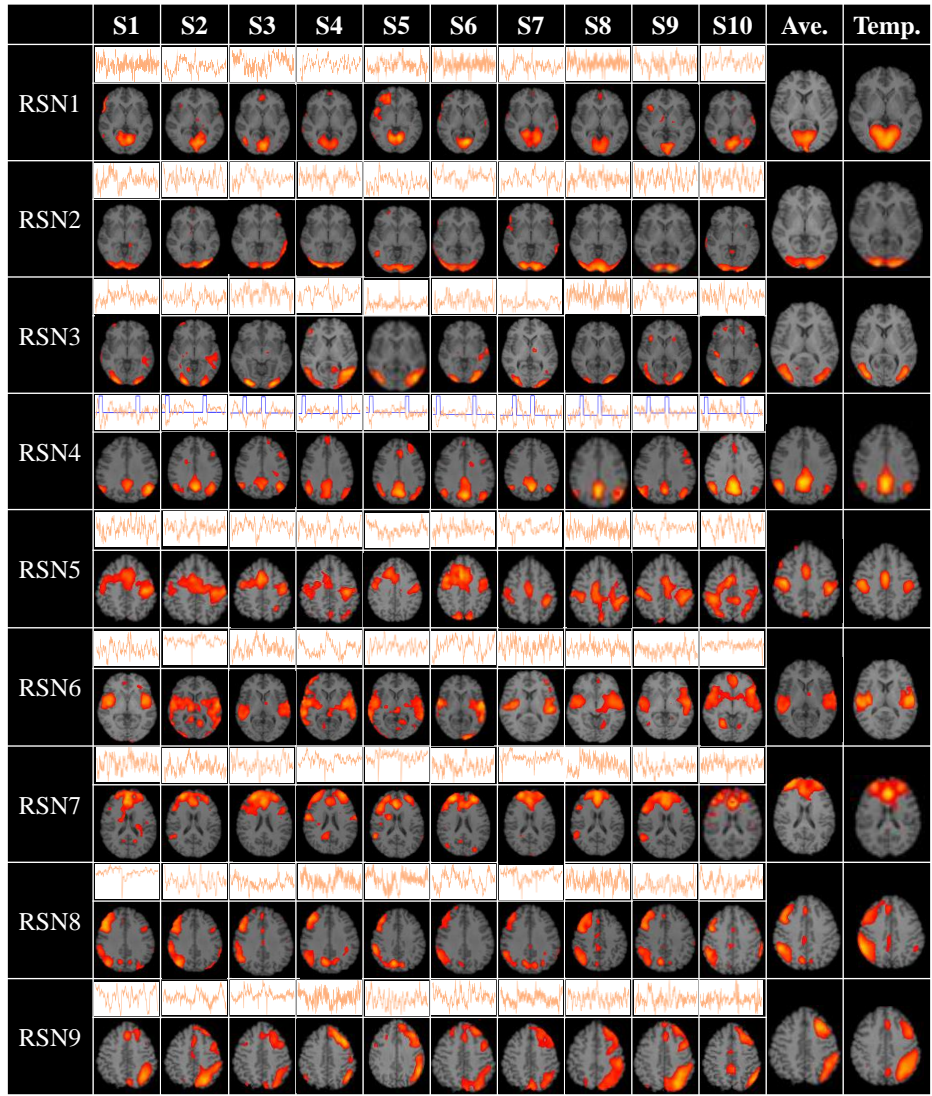